# Modelling and analysis of time in-homogeneous recurrent event processes in a heterogeneous population: A case study of HRTs

Madhuchhanda Bhattacharjee[1,2*]    Elja Arjas[2,3]

1 School of Mathematics and Statistics, University of Hyderabad, India
2 Department of Mathematics and Statistics, University of Helsinki, Finland
3 Department of Biostatistics, University of Oslo, Norway

*: Corresponding author,
email: mbsm@uohyd.ernet.in

Running title: Modelling recurrent events in heterogeneous population

**Abstract:**

In this work we present a method for the statistical analysis of continually monitored data arising in a recurrent diseases problem. The model enables individual level inference in the presence of time transience and population heterogeneity. This is achieved by applying Bayesian hierarchical modelling, where marked point processes are used as descriptions of the individual data, with latent variables providing a means of modelling long range dependence and transience over time. In addition to providing a sound probabilistic formulation of a rather complex data set, the proposed method is also successful in prediction of future outcomes. Computational difficulties arising from the analytic intractability of this Bayesian model were solved by implementing the method into the BUGS software and using standard computational facilities.

We illustrate this approach by an analysis of a data set on hormone replacement therapies (HRTs). The data contain, in the form of diaries on bleeding patterns maintained by individual patients, detailed information on how they responded to different HRTs. The proposed model is able to capture the essential features of these treatments as well as provide realistic individual level predictions on the future bleeding patterns.

# 1 Introduction

Time in-homogeneous data, in particular transience, are frequently encountered in repeated events contexts arising in clinical, biomedical and engineering research problems. Such applications often require further attention to modelling or explaining of cross-sectional heterogeneity due to known or unknown background factors/covariates. Box-Steffensmeier and De-Boef (2002) carried out a Monte Carlo comparison of several event history models. They concluded that the models that do well under heterogeneity don't necessarily do well with added event dependence, and similarly models that do well under event dependence do not also handle heterogeneity well. However it has also been shown that inference quality could suffer if either between subject or within subject heterogeneity is not accounted for (Kaplan et al 2009). Considering both types of dependence jointly presents new challenges for analysts (Torá-Rocamora et al 2013, Bijwaard 2011, Box-Steffensmeier and De-Boef 2006).

There is an extensive literature available on recurrent event data, (ref. Cox 1963, Ascher and Feingold 1983). Many model variations based on the Cox proportional hazards model have also been proposed for the analysis of repeated events (Clayton 1994, Lin 1994, Gao & Zhou 1997, Klein & Moeschberger 1997, Therneau & Hamilton 1997, Wei & Glidden 1997, Hosmer & Lemeshow 1999, Kelly & Lim 2000, Box-Steffensmeir & Zorn 2002). Most of these fall short in capturing the individual/subject level features (in absence of usable covariates) in such problems. Keeping in view today's demand to aid personalized medicine, our objective would be not only to model such recurrent data, in presence of transience and heterogeneity, but also to reliably estimate future outcome trajectories to aid medical decision making.

In the following we focus on description of the data to highlight the complexities involved in the problem. Here we consider data arising from the field of clinical medicine, where the subjects experience repeated relapse of the event. However, at least in part due to the treatment which they are actually undergoing, their condition is expected to gradually improve over time. We propose a model to analyse bleeding pattern of different HRTs, as described by the bleeding diaries maintained by individual women receiving the treatment.

The biological event of menopause signifies an important change in every woman's life.

As menopause approaches, a woman's ovaries start producing less of the female hormones - oestrogen and progestogen. Conditions like osteoporosis develop after menopause because women lose the protective effect of oestrogen. Hormone replacement therapy (HRT) is a medication where oestrogens or a combination of oestrogens and progestogens is applied for the treatment of climacteric symptoms and, for example, for the prevention of osteoporosis. HRTs are known to lead to a significant reduction in the incidence of hip and forearm fractures.

HRT is now available in many different forms and variations. In order to prevent endometrial cancer, women with a uterus are advised to combine oestrogen replacement therapy with one of the many different progestogens available, either in a continuous daily regimen or sequentially, in a monthly or (more experimentally) quarterly regimen. Despite the benefits of HRT to postmenopausal women, a large proportion of the women who start HRT discontinue the therapy after a year, and by three years most patients have done so, due to the occurrence of unwanted uterine bleeding (Hahn 1989, Nachtigall 1993, Hammond 1994, Goldman et al 1998).

The bleeding pattern needs to be studied for two major reasons: firstly to understand the course and outcome of the therapy itself, and secondly to improve patient adherence. Since unwanted bleeding accounts for a high percentage of adherence failures, it is necessary for health care professionals to familiarize themselves with these bleeding data and their inherent pattern.

Bleeding develops regardless of whether the combination regimen used is sequential (Sequential combined HRT: scHRT), that usually produces a regular and therefore predictable bleeding pattern, or continuous administration of the two hormones (known as Continuous combined HRT: ccHRT), producing a much less predictable bleeding pattern. In order to address this common concern, it is necessary for clinicians to decipher the uterine bleeding data on as many HRT products as possible, to be able to guide and assure the patients regarding the course and outcome of the treatment which they receive (Ravnikar 1987, Kadri 1990, Wren and Brown 1991, Ryan et al 1992, Barentsen et al 1993).

Of the different types of HRT, scHRT's could be the less preferred option as they induce a menstrual like withdrawal bleeding occurring towards the end of, or not long after, the progestogen phase. Also in scHRT, bleeding lessens after several years of treatment, and hence many postmenopausal women are unwilling to accept such bleeding episodes for such an extended period of time.

ccHRTs were developed as an alternative to sequential treatments, to improve patient adherence by improving bleeding control. Continuous administration of progestogens or other compounds in combination with oestrogen has been advocated as a means to avoid the scheduled withdrawal bleedings associated with scHRT regimens. ccHRTs generally lead to less frequent and shorter bleeding episodes, but unlike in scHRT, they are not periodical.

The effect of different HRT regimes can be also characterised in terms of lack of bleeding. There are several measures to quantify such developments, most of which are characteristics only at individual time points. In order to predict the chances of bleeding cessation with continued HRT use, summary measurements need to be formed which describe effects over prolonged treatment intervals. A parameter that has been used frequently in recent years is "Cumulative amenorrhea", defined as absence of bleeding or spotting for a continuous period of time, say three or six months.

## 2 Preliminary Exploration of Data

Bleeding information is usually collected in the form of diaries, each diary recording bleeding incidents during 90 consecutive days. Here we consider a data set containing information on three different HRTs, two of which are continuous combined (i.e. ccHRT) whereas the third one is sequential (i.e. scHRT). Data on 163 subjects were collected for approximately one year (4 diaries). Of these 163 subjects, 54 received 'type-1' ccHRT, 56 received 'type-2' ccHRT, and the remaining 53 were given scHRT.

He it may be noted that there is another form of diary data where the events are recorded in a nested manner within each day. This often gives rise to a multi-level organisation of the data (or events of interest) and requires different methods of analyses (see for example Roesch et al 2010).

Additionally we have data from another experiment carried out five years prior to the above trial, where the type-1 ccHRT was applied on 48 subjects and data were collected for two years (referred to as 'second data set' henceforward). We use this latter data set for validation and prediction purposes. In both data sets, while most subjects completed the studies, for a few the treatment and follow-up study had to be terminated due to various reasons.

Next we observe that the nature of bleeding can vary in magnitude. The common practice has been to dichotomise the level into two categories, called "Bleeding" and "Spotting", where spotting is the lesser in discomfort of the two. In the diaries, each date was marked with letter "B" if the patient experienced bleeding, with "S" if she experienced spotting, and it was left blank otherwise (which will be marked by "N" in this analysis).

Depending on the focus of interest, HRT related data can be presented and summarised in a number of different ways. For example, one can consider, for each subject, the percentage of days with bleeding, spotting, and neither bleeding nor spotting. As an alternative, one can consider, at a given time point, the number, or proportion, of subjects with bleeding, spotting and neither, etc. We computed for the second and smaller data set the proportions of days spent by each individual in each of the respective states "B", "S" and "N". As expected, with an exception of a few, the majority of the subjects spent more than 80 percent of their time in state "N", i.e. they were free from both bleeding and spotting (result not shown).

However, it may well be the case that different subjects report comparable values for summary statistics, but then a closer inspection reveals that the number of corresponding episodes is markedly different. In the following the percentage of time spent in the states "B", "S" and "N", along with the total number of such episodes is presented for three (randomly) selected subjects from the same data set (Table 1). The durations of the episodes spent in these three states, for these same subjects, are presented in Figure 1.

(Table 1 approximately here)

(Figure-1 approximately here)

This suggests that every time a subject enters a state not only the type of the state, i.e. "B", "S" or "N", but also the duration in that state as well as the order in which they occur, should be considered for a proper understanding of the course of the treatment.

As was mentioned before, for patients who develop uterine bleeding after initiation of HRT, data on cumulative amenorrhea is thought play an important role in enabling a physician to predict the chances of bleeding cessation with continued HRT use. Thus, cumulative amenorrhea can provide the clinician with valuable information that can be used to help patients to select the HRT best suited to their needs. In order to understand the situation better we carried out a preliminary analysis of the bleeding/spotting free "N" episodes.

In Figure 2 we present several summaries of the "N" episode from the second data set. Top-left panel: This plot shows that the number of "N" episodes experienced by patients varied widely. Top-right panel: The number of "N" episodes varied irrespective of total length of "N" days, (e.g. one patient experienced no bleeding or spotting during the entire study period of 720 days, whereas another patient experienced 46 "N" episodes within a total of 530 "N" days). Bottom-left panel: For any given subject the "N" episode lengths varied highly, with several patients experiencing maximum episode length of more than few hundred days as well as minimum of only a few days. Bottom-right panel: For many subjects there was no apparent trend in the "N" episodes.

(Figure-2 approximately here)

The top panel plots indicate that there is a high degree of heterogeneity across subjects. The bottom panel shows that there can be large variation in the duration of the "N" episodes within a subject, and there is no obvious trend over time. However, the treatment is known to have long term benefits and patients are generally expected to experience gradually longer "N" episodes. It would therefore seem that the speed of improvement is different from patient to patient and needs to be monitored and modelled individually to understand the underlying pattern (Fig. 3).

(Figure-3 approximately here)

## 3 Model and analysis

From the preceding discussion it is apparent that in order to provide a realistic description of the bleeding diary data, and also to lead to useful individual level predictions, the statistical model should respect the following desiderata:
- it should be able to describe adequately the alternating transitions between the non-

bleeding and the bleeding and/or spotting phases;

- it should have the ability to adjust itself to the widely different ways in which individual women respond to HRT, also accommodating the large amount of variation that is seen in the duration of the bleeding and non-bleeding episodes even for the same woman;

- it should have the potential of describing the expected development towards cumulative amenorrhea, but allowing for the fact that this may happen individually at very different speeds;

- given these desiderata and the limited amounts of data, the model structure should be as simple as possible, to enable the estimation of its parameters with a reasonable accuracy, as well as provide realistic individual level predictions regarding the cessation of bleeding.

Our first choice towards a simple model structure concerns the way in which the episodes of non-bleeding, bleeding and spotting registered in the diaries are dealt with. Instead of considering a three state model in which the states would correspond directly to the non-bleeding, bleeding and spotting episodes, we lump the possibly several consecutive episodes of bleeding and spotting that may occur between two adjacent non-bleeding episodes into a single combined episode. In other words, we consider the episode starting from the time at which a previous non-bleeding episode ends, and either bleeding or spotting begins, until the woman enters again the non-bleeding state. However, in order to arrive at a sufficiently fine description of this process, we divide such a combined response further into three types: in a "B" episode there is bleeding, but no spotting, between two adjacent non-bleeding ("N") periods, in an "S" episode there is only spotting but no bleeding, and in a "BS" episode there are one or more consecutive spells of both bleeding and spotting without an intermittent non-bleeding period. (Note that for every completely observed duration in the "lumped B/S/BS" state it will ultimately be clear how it should be classified further according to the finer "B"-, "S"- and "BS"-scale, but from right censored data it may be impossible to distinguish between the "B"- and "BS"-categories on the one hand, and between "S"- and "BS"-categories on the other.)

Another natural thought towards satisfying the above desiderata is to employ a hierarchical model structure where, in addition to the observed alternating durations in the "N" and "B/S/BS" states, there is a latent process that would, in a sense, correspond to the woman's "physiological state" at any given time. The individually different responses over time to the HRT could then be attributed to individually different sample paths of this underlying process. Although, in principle, it would be natural to consider such an underlying state as a continuous variable, it seems that considering $k$ such states, and ordering them in an appropriate way, will be sufficient for an adequate enough description. In this particular example we have used $k = 4$, where one state which (on the observed level) has the tendency of leading to short durations in the "N" phase, another state in which the durations in the "N" phase are (stochastically) somewhat longer, a third one in which they are longer still, and finally a fourth one in which "N" is "nearly absorbing". To roughly approximate an underlying continuous behaviour through a finite number of states, perhaps 3 states in general would be enough, viz. short, medium and long duration. However, for the present data where extreme heterogeneous nature of the data raises the possibility of observing a really long duration equivalent to no relapse to B/S/BS states, necessitates introduction of at least another state, thus leading to the choice of $k = 4$.

In the following we denote these states by N4, N3, N2 and N1, respectively, with N1 being the "nearly absorbing" state. For a woman, whose diary initially documents short durations in the "N"-state, in-between "B"-, "S"- or "BS"-episodes, and is then progressively moving towards amenorrhea, we would expect the underlying state process to move similarly from N4 to N1. Apart from providing a description of such transience in the response of an individual woman to the HRT received, this combination of an observed and an underlying process can be expected to act in the role of a "memory" capturing the essential features of the past recorded history in the diary, and thereby acting as a convenient summary on which the future dynamics can be conditioned.

A natural suggestion, based on the considerations above, would now be to describe the status of an individual woman at time $t$ (when measured from the beginning of the treatment) by a pair of random variables, say $(O_t, L_t)$. Here $O_t$ is the observed status determined from the available diary data and assuming one of the four possible values "N", "B", "S" and "BS", and $L_t$ is the corresponding unobserved (latent) status variable with the possible values 1, 2, 3 and 4 as described above. Recall that the observed process $(O_t)$ is alternating in the sense that after every two transitions it has to be back in state "N". In order to arrive at a sufficiently simple probabilistic description of the bivariate process $(O_t, L_t)$ we now postulate that transitions in the latent process $(L_t)$ can only happen at times at which the process $(O_t)$ returns to the state "N". It would then be attractive to postulate further that the bivariate process $(O_t, L_t)$ has a hidden Markov structure, with $(L_t)$ as the "hidden layer". Unfortunately, already a preliminary analysis of the diary data showed that this would not be realistic, in the sense that even with the latent variable $(L_t)$ included as a part of the "current state" description its knowledge would not provide an adequate basis for predicting the future behaviour of the observed process $(O_t)$. Therefore we need to extend the "memory" from the current state of the process $(O_t, L_t)$ to include also some relevant aspect of its history.

Although not necessarily Markov, $(O_t, L_t)$ nevertheless has the structure of a bivariate jump process, or marked point process, and its probability distribution can be specified in terms of conditional mark specific intensities for the "next" transition, always conditioned on the past history of that process. Because of its "jump-like" character, we can specify such a model either by relating the intensities directly to the follow-up time $t$ of the considered woman, or by indexing the episodes according to their natural ordering in time, corresponding to observed levels of constancy in the process $(O_t)$, and then also taking into account the durations of such episodes together with the relevant status information. Thus, if $0 = T_0 < T_1 < T_2 < \ldots$ are the times at which the observed status $O_t$ of an individual woman changes, we specify a sequence $X_1, X_2, \ldots$ of random durations by $X_j = T_j - T_{j-1}, j \geq 1$.

We also change somewhat the way in which the status was described, that is, by jointly considering $O_t$ and $L_t$. First, we define a corresponding sequence $I_1, I_2, \ldots$ of "phase indicators" by letting $I_j = 1$ if the j-th episode was in an "N" state, that is, if $O_t =$ "N" for $t \varepsilon [T_{j-1}, T_j)$, and otherwise let $I_j = 0$. We then define the sequence $C_1, C_2, \ldots$ by setting

$$C_j = \begin{cases} O_{T_{j-1}} & if\ I_j = 0 \\ L_{T_{j-1}} & if\ I_j = 1 \end{cases}.$$

Note that, since the processes $(O_t)$ and $(L_t)$ always remain constant on intervals of the form $[T_{j-1}, T_j)$, we can say that the variables in the sequence $(C_j)$ alternate between defining an observed or a latent state: during an "N" episode it is identical to the corresponding latent status, and during a bleeding and/or spotting it specifies the type ("B", "S" or "BS") of this episode.

The remaining task is to specify the conditional intensities governing these processes. In order to simplify matters, we postulate that all such intensities are constant over the corresponding episodes, thus giving rise to conditionally independent exponential durations in the corresponding states. Taking into account the earlier remarks on the necessity of including some additional "memory", in order to describe the non-Markovian character of these processes, we have the following two situations to consider:

(i) If the woman is currently, at time $t \in [T_{j-1}, T_j)$, in an "N" state, so that the corresponding phase indicator then has the value $I_j = 1$, we assume that the intensity of leaving that state can depend on the corresponding (latent) state $C_j$. Furthermore, when this transition into the new state $C_{j+1}$ (being then one of the states "B", "S" and "BS") actually happens, we allow the transition probabilities to depend, in addition to the current (latent) status given by $C_j$, also on the (observed) status $C_{j-1}$ during the immediately preceding bleeding and/or spotting episode.

(ii) If she is currently, at time $t \in [T_{j-1}, T_j)$, in one of the observed states "B", "S" and "BS", so that the corresponding phase indicator then has the value $I_j = 0$, we assume that the intensity of leaving that state can depend on the current (observed) state $C_j$. Furthermore, when this transition into the new state $C_{j+1}$ (being then one of the latent states 1, 2, 3 and 4) actually happens, we allow the transition probabilities to depend, in addition to the current (observed) status given by $C_j$, also on the corresponding (latent) status given by $C_{j-1}$.

Thus we could say that this model assumes a "second order memory" structure. Given these as guidelines, we now specify a corresponding hierarchical Bayesian model for the bleeding diary data. In doing so, we need to specify also prior distributions for the necessary model (hyper)parameters. In addition, we have to account for the fact that there were altogether 163 women (now indexed by subscript $i$) with individually different observed and latent status histories, and three different types of treatment (here indexed by symbol $Tr$).

3.1 Model summary:

Here is a summary of the model specification:

1. Duration $j$ of subject $i$:
   $(X_{i,j} \mid Tr_i, I_{i,j}, C_{i,j}, \beta(Tr_i, I_{i,j}, C_{i,j})) \sim Exp(\beta(Tr_i, I_{i,j}, C_{i,j}))$, where $i = 1, \ldots, 163$ and $j \geq 1$,

2. Therapy variables (observed):
   $Tr_i = 1, 2, 3$ indicating type of HRT given to subject $i$

3. Phase variables:
   $I_{i,1} | Tr_i \sim Bernoulli(P_{00}(Tr_i))$ and
   $I_{i,j} = 1 - I_{i,j-1}$ where $i = 1, \ldots, 163$ and $j \geq 2$,

4. Partially latent state variables:
   $(C_{i,1} | Tr_i, I_{i,1}, P_0) \sim Multinomial(1, P_0(Tr_i, I_{i,1}))$,
   $(C_{i,2} | Tr_i, I_{i,2}, P_0) \sim Multinomial(1, P_0(Tr_i, I_{i,2}))$,
   $(C_{i,j} | Tr_i, I_{i,j}, C_{i,j-1}, C_{i,j-2}, P) \sim Multinomial(1, P(Tr_i, I_{i,j}, C_{i,j-1}, C_{i,j-2}))$,
   where $i = 1, \ldots, 163$ and $j \geq 3$,

5. State specific intensities:
   $\beta_{1,0,l} = \beta_{2,0,l} = \beta_{3,0,l} \sim Gamma(0.1, 0.1)$, where $l = 1, 2, 3$ and
   $\beta_{tr,1,1} \equiv 0.00001$,
   $\beta_{tr,1,l} \sim Gamma(0.1, 0.1)$, where $tr = 1, 2, 3, l = 2, \ldots, k$ and
   $\beta_{tr,1,1} \leq \beta_{tr,1,2} \leq \ldots \leq \beta_{tr,1,k-1} \leq \beta_{tr,1,k}$,

6. Transition probabilities:
   $P(tr, 0, l_1, l_2; .) \sim Dirichlet(1_0)$, where $tr = 1, 2, 3, l_1 = 1, \ldots, k$, and $l_2 = 1, 2, 3$,
   $P(tr, 1, l_1, l_2; .) \sim Dirichlet(1_1)$, where $tr = 1, 2, 3, l_1 = 1, 2, 3$, and $l_2 = 1, \ldots, k$,

7. Initial distribution:
   $P_0(tr, 0; .) \sim Dirichlet(1_0)$
   $P_0(tr, 1; .) \sim Dirichlet(1_1)$
   where $tr = 1, 2, 3, 1_0 = (1)_{3\times 1}$ and $1_1 = (1)_{k\times 1}$,

8. Phase probabilities:
   $P_{00}(tr; .) \sim Uniform(0, 1)$ where $tr = 1, 2, 3$.

.

The fixed parameters in the model were chosen so as to give rise to reasonably vague priors. The numerical estimation of the desired parameters and other quantities was carried out using Markov Chain Monte Carlo (MCMC) method, by drawing a large MCMC sample from the posterior distributions of the parameters. The model was specified in BUGS language and implemented in WINBUGS.

**4 Results:**

4.1 Estimation of the structural and population parameters:

The characterising differences between the two types of HRT, viz. continuous combined (cc) and sequential combined (sc), are already visible in the estimated initial probabilities for being in an "N" or a non-"N" state (Table 2) The differences between the two types of ccHRT were negligible.

However, as for scHRT bleeding or spotting occurs mostly around the time of progestine application, initially most subjects (in the present data 98%) could be predicted to be in the "N" phase.

(Table 2 approximately here)

At the initiation of the ccHRT the subjects in the non-"N" phase mostly start with spotting, i.e. in the "S" state (Table 3). A closer inspection reveals the heterogeneous nature of the study sample at the beginning of the treatment. The estimates for the scHRT appeared to be well in line with the existing knowledge of the behaviour of this type of therapy, which induces a rather regular and menstrual like response in the subjects.

(Table 3 approximately here)

So far, based on the initial phase and latent state distributions, the two ccHRTs had not shown any noticeable difference. However, the intensity estimates (Table 3) brought out some interesting differences between the two. Subjects receiving Type-1 ccHRT would be expected to experience much shorter bleeding episodes than those receiving Type-2 ccHRT (the respective intensity estimates of leaving the S-state being 0.75 and 0.29.) . The pattern of estimates for scHRT was again different from those for ccHRTs, the most noticeable difference being in the expected duration of the "SB" episodes.

In the following the estimates of the two-step transition matrices are presented for all three treatments and two phases (Table 4). Recall that even though the patient's condition is generally expected to improve over time, a preliminary examination of the data had not shown any readily recognizable trend in this respect. Therefore, due to lack of evident trend over time, we did not impose any a priori monotonicity condition on the transition matrix. We had also assumed vague priors on the transition probability matrices. Nevertheless, the posteriors show noticeable patterns.

(Table 4 approximately here)

Of the two ccHRTs, the transient nature of type-1 ccHRT can be seen clearly. For example, a woman initially in the latent non-bleeding state N4 could be expected, with probability 0.31, to be in a "better" state N2 after having subsequently again returned to a non-bleeding phase, and then with probability 0.35 in state N1 after the next return. For type-2 ccHRT, the chance of remaining in an initial state N4 seemed somewhat higher than for type-1. Otherwise, however, the transient behaviour of moving towards the better states was even clearer in type-2 .

Based on the medical background behind scHRT, subjects undergoing this therapy are expected to experience episodes in state "N" of approximately the same length each and improvement only over long term. Note that the estimates also predict the same, although no such a priori assumptions or restrictions were imposed on the relevant parameters.

The two ccHRTs show noticeably different trends for the bleeding and spotting episodes. Type-1 ccHRT shows much higher probabilities for transition to a bleeding episode, compared to those under type-2 ccHRT. However, recall that the bleeding episodes under type-1 ccHRT were predicted to be very short since the intensity estimate of leaving that state was much larger than all the others.

For the scHRT, irrespective of the current "S/SB/B" episode type there is high possibility for transiting to an "SB" state in the next cycle. However, the estimated intensity associated with "SB" episodes for subjects under scHRT was also comparatively higher, indicating that the expected durations of the episodes in that state are shorter than those experienced under ccHRTs.

4.2   Prediction for generic individuals:

In order to provide useful summaries of the estimated complex model, we computed predictive distributions for the event processes for generic individuals assumed to undergo different HRTs. The posterior means of the model parameters presented above in the Tables 2, 3 and 4 were used to simulate these processes. The prediction period was the first year of the therapy. The simulated processes can be summarised in several different ways. In the following we present the corresponding predictive probabilities for being in state "N" for a generic individual, considered at different time points during this one year.

Although initial data explorations had not revealed any systematic trend of moving towards amenorrhea, the predictions for a generic individual clearly indicated this to be the case (Figure-4). This became evident already from the parameter estimates for type-1 and type-2 ccHRT. Subjects receiving ccHRT were predicted to improve quite fast over the one year period, whereas corresponding predictions for scHRT show no, or only a very slow trend.

(Figure-4 approximately here)

As mentioned before, the effects of different HRT regimes can be also characterised in terms of lack of bleeding. In order to predict the chances of bleeding cessation with continued HRT use, measures need to be used which describe effects over prolonged treatment intervals. For the purpose of analysis, cumulative amenorrhea was defined to be absence of any bleeding or spotting for a period of 180 consecutive days (approximately six months), and similarly the waiting time to cumulative amenorrhea was defined as the time to last bleeding or spotting that is then followed by absence of it for more than 180 days.

Table 5 provides a brief summary of the results that were obtained, in terms of the median and the mean of the predictive distribution for the waiting time to cumulative amenorrhea. The results also showed that, once cumulative amenorrhea is achieved it is also usually sustained for quite long.

(Table 5 approximately here)

We also investigated, not only the time till cumulative amenorrhea is achieved, but the nature of progression during the waiting time under the different HRTs. For this purpose, the simulated processes were used to predict the percentage of time spent in the different states, i.e. the proportions spent in different latent "N" states and in the "S/SB/B" states (see Table 5 for summary results).

Predictive distributions for the residual time to the next relapse at pre-determined time points and for different subjects were also computed. Judged from these predictive distributions, the residual times appear to be stochastically ordered. In other words, the more time has elapsed from the beginning of the treatment, the longer, in the sense of stochastic ordering, it takes to the next relapse (results not shown).

4.3  Prediction for real individuals:

Note that the above predictions for generic individuals can also be thought of as predictions for a real individual at the beginning of the therapy when there are no individual covariate or follow-up data available. Once a subject starts to undergo the therapy, the resulting information on the progression of the treatment outcome can be used to predict her future response to the treatment. This can be done as the proposed model has the capacity to storage into its memory some key features of the earlier response of that same subject, both observed and latent, and then use this information in the predictions.

Here we consider the second data set, wherein follow-up information was collected for two years. Using observations from the first year (of the two years' of data) for 5 selected individuals, and the results from our analysis of type-1 ccHRT based on data set-1, we made an attempt to predict how these 5 subjects would respond to the treatment after the first year. The predictions were then compared with the actual observed data from the second year for the same subject. The "N" episodes from the full two years' data for these five subjects are presented in Figure 5. Note the distinctly different response to the therapy of these five subjects undergoing the same HRT.

(Figure-5 approximately here)

For each of these five individuals we computed the predictive probabilities for being in state "N" at different time points in the second year (Figure 6). Observe from the first year's data that, apart from subject 4, the others had experienced several "N" episodes of a short duration. However, the model is able to identify the characteristic differences between the subjects quite well, with the consequence that the predictive probabilities differ from individual to individual.

(Figure-6 approximately here)

Note in particular that the improvement observed in subject 2 in the second year is sudden and entirely different in nature than the "N" episodes experienced by the this subject in the first year. However, the model successfully estimated the latent state of the subject and gives a similar predictive probability for being in state "N" as to subject 4, who had responded well to the therapy during its entire course.

Also note that subject 3, who had initially shown improved response to the HRT with longer "N" episodes, gradually experienced worsening in her condition and several short episodes. The model gives her the lowest predictive probabilities for "N" episodes among all five.

## 5 Discussion

We feel that our statistical model, combined with Bayesian inferential and computational methods in the estimation of its parameters, fulfils rather well the desiderata presented at the beginning of Section 3. In particular, the example considered above in Section 4.3 demonstrates that the method can provide realistic prognoses for an individual patient, when diary information from monitoring the condition of this patient is combined with corresponding background sample data. Such knowledge has some potential of being usefully applied in clinical practice, in which case it could be seen as a form of personalized medicine. On the other hand, we acknowledge that the computational burden in applying such a method is sizeable, which at present is likely to prevent its routine application for such a purpose.

There have been few methodological developments in this particular area in the recent literature. In related areas we have seen routine implementations of existing software (e.g. Gerlinger et al 2009), which however are unable to capture the complex heterogeneous structure of the problem. In the context of other applications, some efforts have been made to address similar problems in the classical (e.g. mazroui et al 2013) or in the Bayesian framework (e.g. Ding et al 2012, Pennell et al 2010).

While the construction of the model presented here was tailored to fit to considering HRTs and the particular data which were available to us, the above desiderata represent features which we would expect to be characteristic to numerous other interesting applications, and then shared in the corresponding data sets. These features include alternation, or more general changes, between two or more observable states for each considered individual or item, progressive development in its condition and corresponding transience in the sojourns in such states over time, non-Markovian dependence with respect to the observable states, and a high degree of heterogeneity of the individuals or items in the considered population.

Among potential applications of this type one could include, for example, data collected from monitoring the use, maintenance and repair of some particular technical devices, data on employment histories with information on sick leaves, hospital stay data for patients with a particular chronic disease, or recidivism of criminal offences and jail sentences. Depending on the application in mind, it may be important to supplement the model with individual covariate data, an aspect which we have ignored in our present study of HRTs.

The answers to the stated concrete questions are often most usefully given in the form of a probabilistic prediction of the future development of the considered process. A very special feature of our suggested approach is that such predictions can then be concerned with a particular individual, or be provided on the level of the considered background population.

**Acknowledgement**

We are grateful to Bayer Oy, Finland for a permission to use the present data set for our statistical analyses. The initial effort of M.B., which later led to this document, was supported by Schering Oy, Finland.

**Figure Legends:**

Figure 1: Graphical presentation of the N (=1), S(=2) and B(=3) episodes (in vertical axes) against each day (in *Y axes) for t*hree selected individuals with comparable summary.

Figure 2: Summaries based on "N" episodes from second data set.

Figure 3: Durations of "N" episodes of some selected individuals from data set-2. The horizontal axes present the number of the "N" episodes only, where the vertical axes represent duration in days.

Figure 4: Predicted probability for "N" state or otherwise at fixed time points for generic individuals undergoing different HRT regimes, with the estimated probability in Y axis and cumulative days under treatment presented in X-axis.

Figure 5: Durations of "N" episodes for five selected individuals from data set 2. The horizontal axes represent the episode number, the vertical axes present the durations of each episode in days, with episodes from the first year marked in black and those from second year are in white.

Figure 6: Predictive probabilities for being in state "N" at different time points in the second year (based on data from first year only) for 5 individuals from data set 2.

.

Table 1: For three randomly picked subjects, the number (and percentage) of B, S and N days, along with the total durations and the number of episodes.

| Subject | N days (%) | S days (%) | B days (%) | Total duration (No. of episodes) |
|---|---|---|---|---|
| A | 648 (90.0) | 58 (8.1) | 14 (1.9) | 706 (69) |
| B | 650 (90.3) | 70 (9.7) | 0 (0.0) | 720 (36) |
| C | 664 (92.2) | 54 (7.5) | 2 (0.3) | 718 (9) |

Table 2: HRT specific posterior means of "phase probabilities" $P_{00}$ for initially being in an "N" state, together with the corresponding conditional probabilities for then being in one of the four latent states (left box), and for initially being in a non-"N" state, with the corresponding conditional probabilities for then being in one of the three alternative observed states (right box).

| HRT | Phase type N | | | | | Phase type S/SB/B | | | |
|---|---|---|---|---|---|---|---|---|---|
| | $P_{00}$ | $P_0$ | | | | $1-P_{00}$ | $P_0$ | | |
| | | 1 | 2 | 3 | 4 | | S | SB | B |
| Type-1 ccHRT | 0.37 | 0.07 | 0.09 | 0.28 | 0.56 | 0.63 | 0.82 | 0.14 | 0.04 |
| Type-2 ccHRT | 0.34 | 0.11 | 0.17 | 0.17 | 0.56 | 0.67 | 0.81 | 0.13 | 0.06 |
| scHRT | 0.98 | 0.02 | 0.66 | 0.31 | 0.01 | 0.02 | 0.18 | 0.76 | 0.06 |

Table 3: HRT-specific posterior means of state specific intensities $\beta_{0,l}$ (for N-phase, $l = 1, \ldots, 4$) and $\beta_{1,l}$ (for non-N-phase, $l = 1, 2, 3$).

| HRT | "N"-states[a] | | | | Non-"N"-states | | |
|---|---|---|---|---|---|---|---|
| | 1 | 2 | 3 | 4 | S | SB | B |
| Type-1 ccHRT | | | | | 0.19 | 0.13 | 0.75 |
| Type-2 ccHRT | 0.00 | 0.02 | 0.04 | 0.10 | 0.24 | 0.16 | 0.29 |
| scHRT | | | | | 0.33 | 0.44 | 0.24 |

a: The intensity is assumed to be same for all three HRTs, see model specification.

Table 4: Posterior means of HRT specific two-step transition probabilities, $P(l, 0, , )$ for the "N"-states (left), and $P(l, 1, , )$ for the "S/SB/B"-states (right).

| State | N1 | N2 | N3 | N4 | State | S | SB | B |
|---|---|---|---|---|---|---|---|---|
| | | | | type-1 ccHRT | | | | |
| N1 | 0.53 | 0.20 | 0.14 | 0.13 | S  | 0.89 | 0.09 | 0.02 |
| N2 | 0.35 | 0.48 | 0.07 | 0.09 | SB | 0.63 | 0.33 | 0.03 |
| N3 | 0.04 | 0.22 | 0.35 | 0.40 | B  | 0.18 | 0.09 | 0.73 |
| N4 | 0.02 | 0.31 | 0.07 | 0.59 | | | | |
| | | | | type-2 ccHRT | | | | |
| N1 | 0.64 | 0.16 | 0.09 | 0.11 | S  | 0.91 | 0.08 | 0.02 |
| N2 | 0.43 | 0.43 | 0.08 | 0.06 | SB | 0.81 | 0.19 | 0.00 |
| N3 | 0.17 | 0.22 | 0.39 | 0.22 | B  | 0.71 | 0.14 | 0.14 |
| N4 | 0.02 | 0.16 | 0.07 | 0.75 | | | | |
| | | | | scHRT | | | | |
| N1 | 0.27 | 0.12 | 0.41 | 0.19 | S  | 0.44 | 0.54 | 0.02 |
| N2 | 0.04 | 0.56 | 0.32 | 0.07 | SB | 0.22 | 0.76 | 0.03 |
| N3 | 0.00 | 0.02 | 0.82 | 0.16 | B  | 0.31 | 0.44 | 0.25 |
| N4 | 0.00 | 0.03 | 0.77 | 0.19 | | | | |

Table 5: Results for a generic individual: HRT specific (posterior) predictive distribution for the time to cumulative amenorrhea, and the predicted proportions of time spent in different states till cumulative amenorrhea.

| HRT | Time to cumulative amenorrhea | | Proportions of time spent | | | | | | |
|---|---|---|---|---|---|---|---|---|---|
| | | | Phase type N | | | | Phase type S/SB/B | | |
| | Median | Mean | 1 | 2 | 3 | 4 | S | SB | B |
| type-1 ccHRT | 99.8 | 153.1 | 0.00 | 0.48 | 0.21 | 0.13 | 0.11 | 0.06 | 0.01 |
| type-2 ccHRT | 75.1 | 121.4 | 0.00 | 0.47 | 0.20 | 0.15 | 0.11 | 0.05 | 0.02 |
| scHRT | 327.7 | 535.7 | 0.00 | 0.23 | 0.56 | 0.07 | 0.03 | 0.04 | 0.01 |

Figure-1

Figure-2:

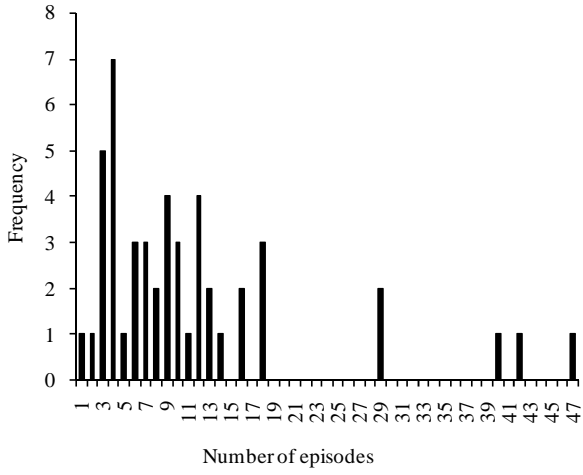
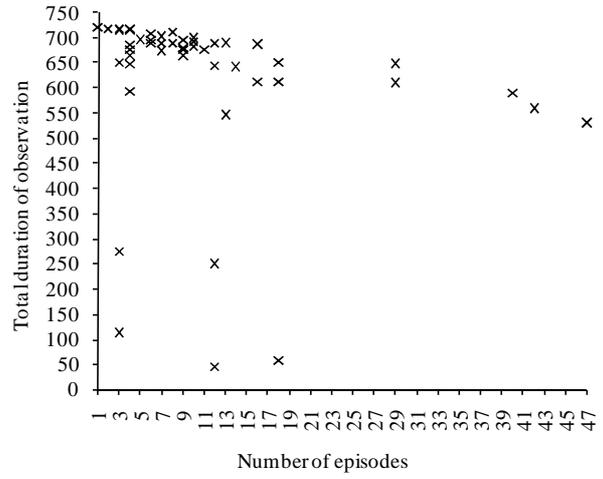
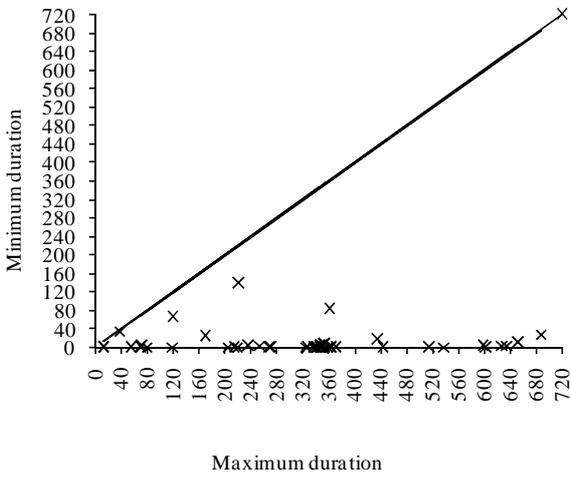
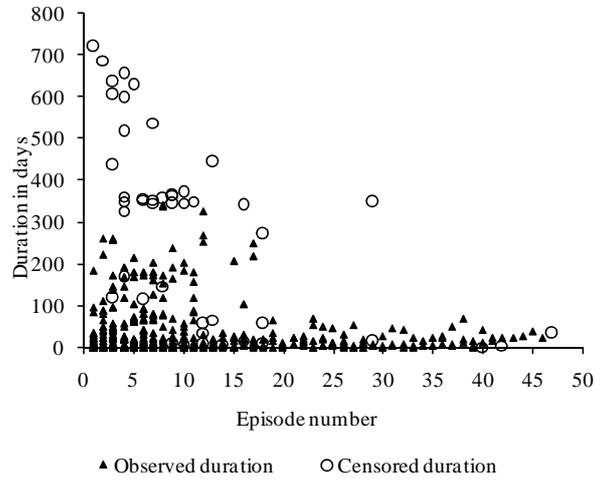

Figure-3:

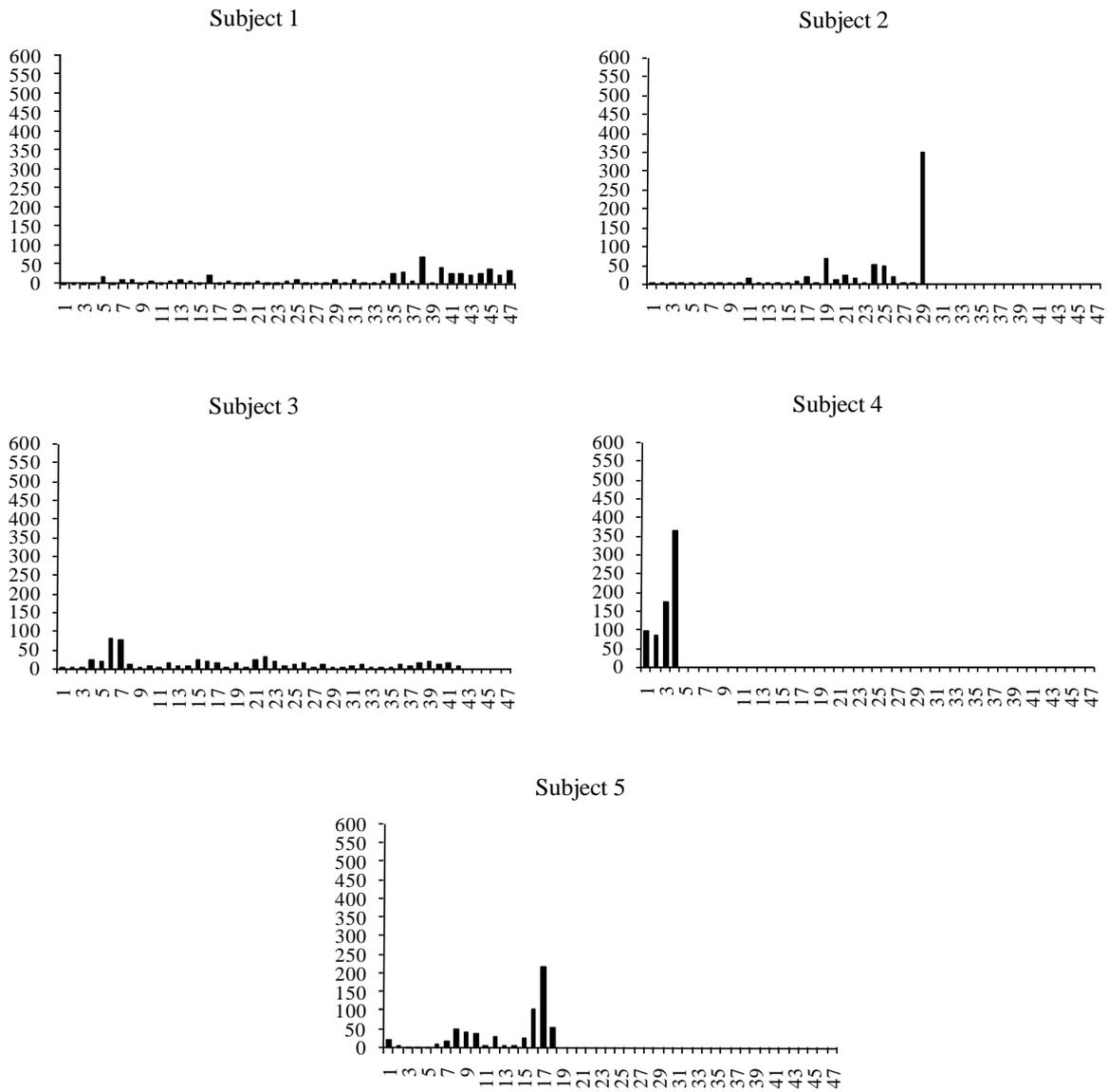

Figure-4:

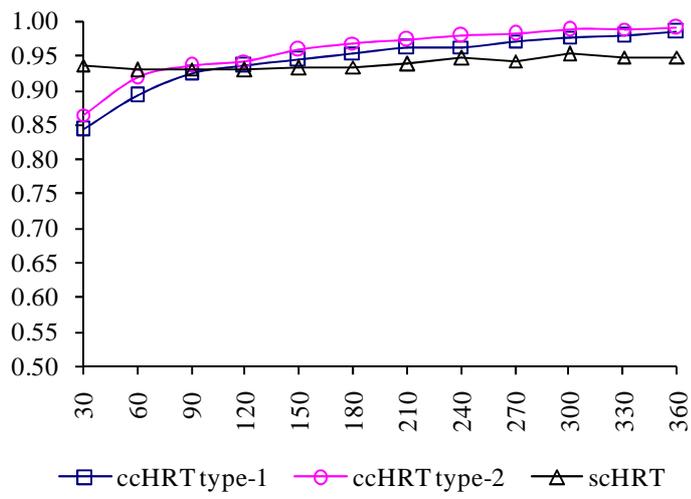

Figure-5:

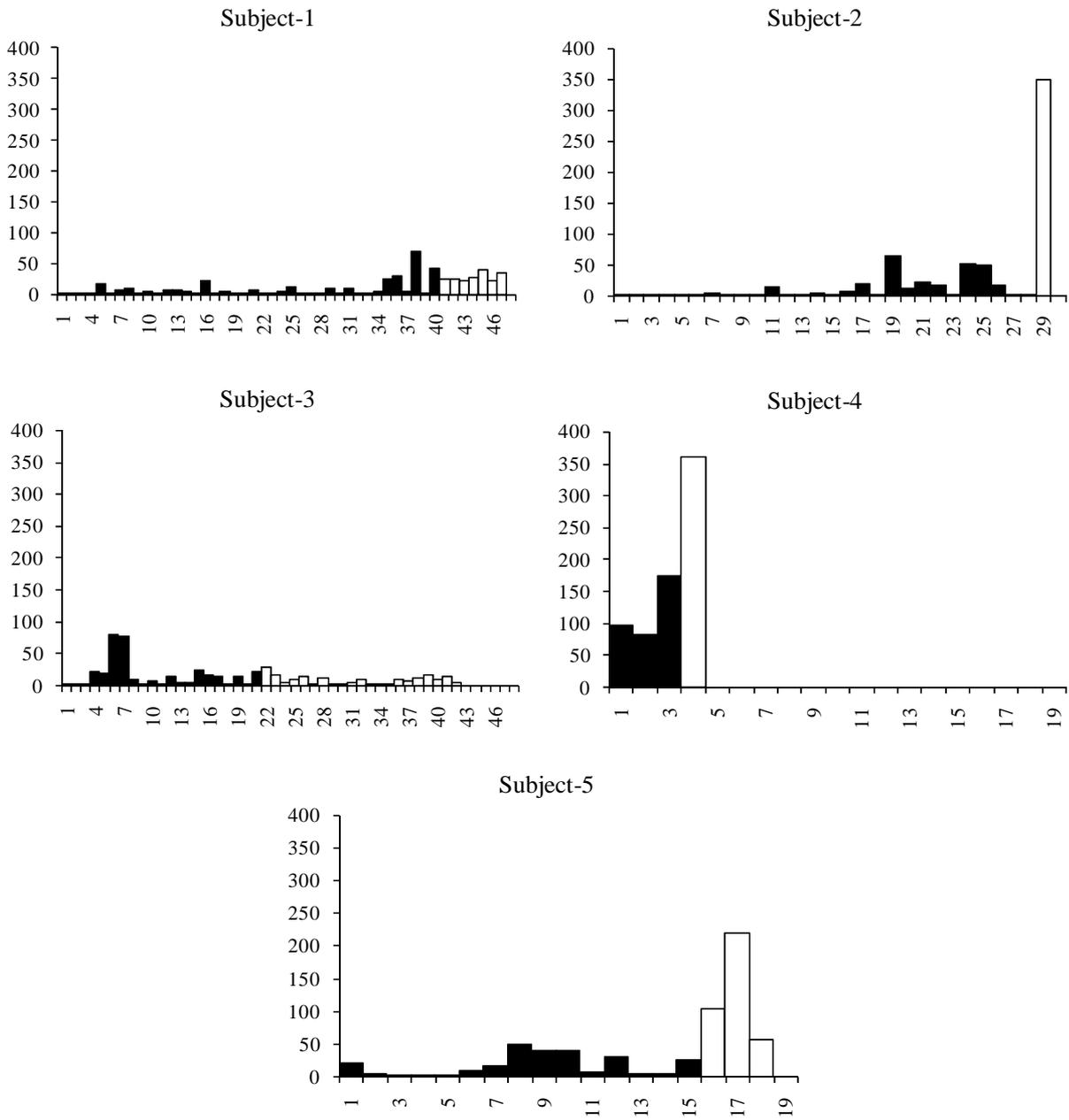

Figure-6:

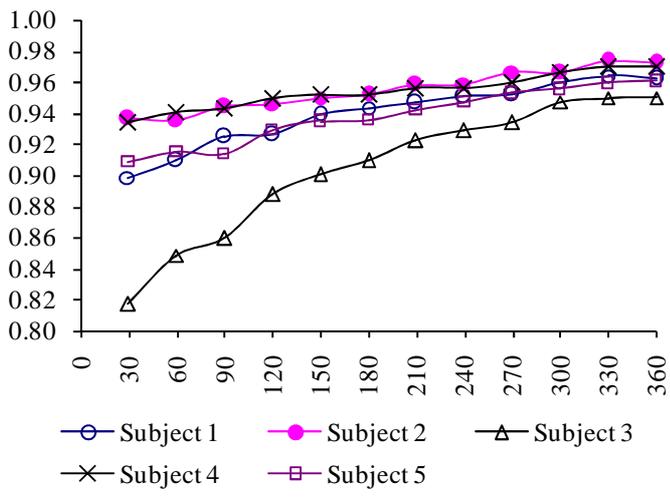